\documentclass[12pt]{article}
\usepackage{cite}
\usepackage{color}
\usepackage{graphicx}
\usepackage{amsmath}
\usepackage{amssymb}
\usepackage{xspace}

\def\baselinestretch{1.2}
\newcommand{\be}{\begin{equation}}
\newcommand{\ee}{\end{equation}}
\newcommand{\beq}{\begin{eqnarray}}
\newcommand{\eeq}{\end{eqnarray}}

\newcommand{\gone}[1]{{}}

\begin{document}
\begin{titlepage}
\begin{flushright}
MAD-TH-08-01
\end{flushright}

\vfil\

\begin{center}

{\Large{\bf A Note on Spontaneously Broken Lorentz Invariance}}

\vfil

Akikazu Hashimoto

\vfil

Department of Physics, University of Wisconsin, Madison, WI 53706

\vfil

\end{center}

\begin{abstract}
\noindent We consider a relativistic effective field theory of vector
boson whose vacuum exhibits spontaneous breaking of Lorentz
invariance. We argue that a simple model of this type, considered
recently by Kraus and Tomboulis, is obstructed from having a
consistent ultraviolet completion according to the diagnostic recently
suggested by Adams, Arkani-Hamed, Dubovsky, Nicolis, and Rattazzi.
\end{abstract}
\vspace{0.5in}

\end{titlepage}
\renewcommand{\baselinestretch}{1.05}  

Relativistic quantum field theory, formally defined as Lorentz
invariant ultraviolet (UV) fixed point and a subsequent
renormalization group flow, is an elegant formalism successfully
capturing the physics of elementary particles.  As a quantum theory,
they rest on the foundation of unitary evolution of states in the
Hilbert space.  States themselves form a representation of the Lorentz
group, and the dynamics dictates the spectrum of states and their
interactions.

Gauge principle is a critical ingredient for keeping unitarity and
Lorentz invariance mutually compatible in models including states of
spin one (or higher).  When combined with the requirement that the UV
fixed point is a weakly coupled field theory with a Lagrangian
formulation, the list of possible models is incredibly short. They
essentially consist solely of asymptotically free gauge theories
minimally coupled to matter fields which are spin zero or spin half.
It is therefore quite remarkable that QCD is, and that the Standard
Model admits a UV completion of,\footnote{An example of embedding of
the Standard Model in a conformal UV fixed point was described in
\cite{Frampton:1999yb}.}  a model of this type.

A closely related and familiar formalism in a theorist's toolbox is
the effective field theory (reviewed, e.g., in
\cite{Polchinski:1992ed,Kaplan:1995uv,Manohar:1996cq}).  Roughly
speaking, effective field theory is the result of flowing down the
renormalization group from the UV to some scale of interest from the
point of view of a probe or a physical process.  It is not possible to
track the flow of all the parameters of the renormalization group in
closed analytic form.  Nonetheless, one can draw useful conclusions
about the strength of physical effects through systematic analysis of
the dimensions of operators in the effective action, energy scales,
and symmetries, within the framework of effective field theory.  A
productive and frequently adopted attitude in the use of effective
field theory technique is to not dwell on the UV completion. By
following this dictum, model builders in cosmology, astrophysics, and
particle physics are free to introduce wide range of exotic models and
to study their implications.

In light of the restrictiveness imposed by the consistency, however,
there always persist some degree of doubt that a given effective field
theory model might {\it not} admit a consistent UV completion. This is
especially the case for exotic models exhibiting features such as the
spontaneous breaking of Lorentz invariance. It is not difficult to
imagine that effective field theories arising from a consistent UV
complete theories are somehow restricted, e.g., in the numerical
values of the coefficients of various operators in the effective
action.  These conditions have not been explored systematically simply
because they are difficult to derive from first principles.

Along this line of thought, a simple connection relating positivity of
forward scattering amplitude, possibility of superluminal propagation
in a non-trivial background, and a sign of certain irrelevant
operators in the effective field theory, was pointed out recently by
Adams, et.al. in \cite{Adams:2006sv}.  A prototype of their argument
is an effective field theory of the form
\be L = \int d^4 x \, \left[{1 \over 2} \partial_\mu \phi \partial^\mu \phi
+ c  (\partial_\mu \phi \partial^\mu \phi)^2 \right]\ . \label{prototype} \ee
We take the space-time to be Minkowski. Since much of the discussion
boils down to that of a sign, let us also specify that we are using
the metric signature convention $(+,-,-,-)$. The authors of
\cite{Adams:2006sv} have argued that the sign of $c$ must be {\it
positive} if such an effective field theory is to arise from a
consistent UV complete theory, by showing that theories with negative
values of $c$ exhibits pathologies which manifests itself in the form
of a superluminal propagation in a non-trivial background. The authors
of \cite{Adams:2006sv} also showed that the negative sign of $c$ is in
conflict with {\it analyticity} of the forward scattering amplitudes. These
findings further highlight the significance of the sign of $c$ as a
way to diagnose whether a certain effective field theory can arise
from a relativistic UV fixed point theory. The authors of
\cite{Adams:2006sv} identified the induced gravity models of Dvali,
Gabadadze, and Porrati \cite{Dvali:2000hr} as an example of a model
which appears not to admit a UV completion from this point of view.

In this article, we investigate the status of a model of spontaneous
Lorentz symmetry breaking along similar lines. More specifically, we
study a model whose effective Lagrangian (so the theory has a cut-off)
has the form\footnote{Effective field theory models of spontaneous breaking of Lorentz invariance have been studied extensively by Kostelecky and collaborators, e.g., in \cite{Kostelecky:1989jw,Bluhm:2004ep,Bluhm:2007bd}.}
\be L = \int d^4 x \,  N \left[ - {1 \over 4} F_{\mu \nu} F^{\mu \nu} - V(A_\mu A^\mu) \right] \label{eff} \ . \ee
The parameter $N$ is a dimensionless constant and is taken to be large
so that the theory is weakly coupled. The vector field interacts via
the Lorentz invariant interaction term $V(A^2)$, which one might take
to be of the form
\be V(A^2) = {\lambda \over 4} (A_\mu A^\mu  - v)^2  \label{Vpot} \ee
which is minimized at
\be A^2 = A_\mu A^\mu = v \ . \ee
This can  a priori can be either positive (time-like) or negative
(space-like) depending on the sign of $v$. More generally, we take $V(A^2)$ to admit a power series expansion in $A^2$ of the form
\be V(A^2) = v_2 A^2 + v_4 (A^2)^2 +  v_6 (A^2)^3 +  \ldots \ , \qquad v_n \sim \Lambda^{4-n} \ . 
\ee
The goal of this note is to argue that (\ref{eff}) is an inconsistent
effective field theory according to the diagnostic criteria of
\cite{Adams:2006sv}.

The general idea that photons might arise as a Goldstone boson for
spontaneously broken Lorentz invariance has a long history and appears
to go back to Bjorken \cite{Bjorken:1963vg} who considered a fermion
model similar to that of Nambu and Jona-Lasinio \cite{Nambu:1961tp}
but with a condensation in the vector component of the fermion
biliniear
\be 
L = \bar \psi_i (i \partial\!\!\!/-m) \psi_i +  \lambda_{2} (\bar \psi_i \gamma^\mu \psi_i)^2  \label{bjorken}
\ . \ee
The original vision of Bjorken was to obtain a Lorentz invariant
dynamics of quantum electrodynamics in a non-Lorentz invariant
gauge. It was later pointed out by Banks and Zaks \cite{Banks:1980rh},
however, that the violation of gauge invariance is physical and has an
effect on gauge invariant observables. More recently, Kraus and
Tomboulis studied an effective field theory precisely of the form
(\ref{eff}) and argued that
\begin{itemize}
\item such an action can arise as an effective dynamics for a model
similar to that of Bjorken where $N$ corresponds to the number of
fermion fields,\footnote{The connection between the fermion model and
the effective vector model was analyzed critically in
\cite{Jenkins:2003hw}.  We thank Per Kraus for bringing this article
to our attention. Independent of this observation, we can explore the
UV completability issue for the effective field theory of vector boson
model (\ref{consider}).} and
\item that the observable effects of Lorentz invariance can be made
parametrically small by taking $N$ to be large.
\end{itemize}

To relate (\ref{eff}) to models of fermion bilinear condensates, Kraus
and Tomboulis considered a model of the form
\be 
L = \bar \psi_i (i \partial\!\!\!/-m) \psi_i +  N \sum_{n=1}^\infty \lambda_{2n} {(\bar \psi_i \gamma^\mu \psi_i)^{2n} \over N^{2n}} \label{KT}
\ , \ee
with the dimensionless coupling $\lambda_{2n}/\Lambda^{4-6n}$ being a
number of order one. An effective field theory of this form might 
arise, for example, by integrating out a massive vector boson of mass
$\Lambda$ with respect to which the fermions are charged. 

The action (\ref{KT}) can be re-written using the standard trick of
introducing an auxiliary field $A^\mu$
\be L = \bar \psi_i (i \partial\!\!\!/-A\!\!\!/-m) \psi_i -  N V(A_\mu A^\mu)
  \ . \label{aux1} \ee
In this formulation,  $A_\mu$ is non-dynamical and imposes a constraint
\be \bar \psi_i \gamma_\mu \psi_i  + 2 N A_\mu V'(A^2) = 0 \ , \ee
and when $A_\mu$ is eliminated, reduces  (\ref{aux1}) to (\ref{KT}) where
\be \lambda_2 = {1 \over 4  v_2}, \qquad \lambda_4 = -{v_4 \over 16 v_2^4}, \qquad \lambda_6 = {4 v_4^2 - v_2 v_6 \over 64 v_2^7}, \qquad \ldots \ . \ee
If instead we integrate the fermions out first, a gauge invariant
kinetic term of order $N$ is induced and we recover an effective
action of the form (\ref{eff}), up to numerical factors of order one
and additional higher derivative operators.

The case where only $v_2$ and $\lambda_2$ are non-vanishing
corresponds to the model of Bjorken. The effective action in terms of
$A_\mu$ field is the Proca action
\be L = \int d^4 x \,  N \left[ - {1 \over 4} F_{\mu \nu} F^{\mu \nu} + {\mu^2 \over 2}  A_\mu A^\mu \right] \label{proca} \ee
where
\be \mu^2 = - 2v_2 = - {1 \over 2 \lambda_2} \ . \label{mulambda2}\ee
The resulting dynamics, however, is crucially sensitive to the sign of
$\lambda_2$ \cite{Bjorken:1963vg,Banks:1980rh}. If $\lambda_2$ is
negative, space-like components of the $A_\mu$ are stable and we
expect to find the usual effective dynamics of massive vector
bosons. This theory is in fact renormalizable and BRST invariant, and
can be quantized perturbatively when coupled minimally to matter
\cite{Collins:1984xc}. If $\lambda_2$ is positive, one expects a run away behavior,
which must somehow be stabilized in order for the theory to have a
vacuum. Higher order operators in $V(A^\mu A_\mu)$ were introduced
specifically for this purpose \cite{Kraus:2002sa,Bjorken:2001pe}.

While the effective Lagrangian (\ref{eff}) does appear to arise from a
model of interacting fermions (\ref{KT}), it does not necessarily
follow that (\ref{eff}) can be embedded into a consistent UV fixed
point theory. The reason is simply the fact that (\ref{KT}) is not a
renormalizable field theory. While an effective action whose general
from is that of (\ref{KT}) could arise from integrating out a massive
vector field, it is not obvious that there are enough freedom to
control the values of the coupling constants $\lambda_{2n}$ in the
effective action. It is unnatural, in particular, for $\lambda_2$ to
take on a positive value as a result of integrating out a massive
vector boson, which corresponds to fermion/anti-fermion interactions
being repulsive.

More can be said about the range of possible values of $\lambda_{2n}$
if we were to consider this model in 1+1 dimensions. There, we can
take advantage of bosonisation techniques (reviewed
e.g. in\cite{Abdalla:1991vua,Frishman:1992mr,Stone:1995ys}) to study these models in greater
detail. The 1+1 dimensional version of (\ref{KT}) is in fact a
generalization of the Thirring model which can be analyzed along the lines of 
\cite{Coleman:1974bu,Mandelstam:1975hb,Naon:1984zp},  with additional
higher order interactions of the form
\be \lambda_{2n} (J_\mu J^\mu)^n \sim \lambda_{2n} \left((\bar \psi \gamma_\mu \psi)(\bar \psi \gamma^\mu \psi)\right)^n  \sim \lambda_{2n} (\partial_\mu \phi \partial^\mu \phi)^n \ee
In terms of the bosonized scalar field, this model resembles the
``Ghost condensation'' model \cite{ArkaniHamed:2003uy} if
$\lambda_2$ is positive, and might also contain an additional
sine-Gordon potential if the fermions are massive.  The coupling
constant $\lambda_4$ is naturally seen in this formulation to be
constrained by the diagnostic of \cite{Adams:2006sv}. It is very
natural to expect the values $\lambda_{2n}$ of the irrelevant fermion
couplings in 3+1 dimensions to be constrained along similar lines,
although we were unable to find a simple formulation of it at the
present time.

These considerations, however, more directly concerns the fermion
model (\ref{KT}) than the effective theory (\ref{eff}). In fact, it is
well known that in 1+1 dimensions, the fermions do not even induce a
kinetic term for the gauge fields. We will therefore shift our focus
away from the fermionic formulation (\ref{KT}) and concentrate on
(\ref{eff}) more directly.

A different way in which (\ref{eff}) can arise as an effective field
theory is from a renormalization group flow of an Abelian Higgs system
\be L = \int d^4 x \,  N \left[ - {1 \over 4} F_{\mu \nu} F^{\mu \nu} - V\left({1 \over  \phi_0^2} D_\mu \phi^* D^\mu \phi \right) + V_\phi(\phi^* \phi) \right]   \label{eff2} \ee
where the Higgs potential $V_\phi$ is minimized at $|\phi| =
\phi_0$, and the covariant derivative 
\be D_\mu \phi = (\partial_\mu + i A_\mu) \phi \ . \ee
The standard Higgs mechanism will give rise to a massive scalar Higgs
and three components of the vector field. The effective action
(\ref{eff}) emerges by integrating out the massive Higgs. If the
potential $V(A^2)$ has the form given in (\ref{Vpot}), we see that
this gives rise to a term in the effective action of the form
\be  L = -{1 \over \phi_0^4} (D_\mu \phi^* D^\mu \phi)^2 \label{higgs} \ee 
with a {\it negative} coefficient. The criteria of \cite{Adams:2006sv}
applies in classifying (\ref{eff2}) as not having a consistent UV
completion.

While these results are suggestive of (\ref{eff}) being inconsistent,
one can still argue that it is a symptom specific to the massive Higgs
field in (\ref{eff2}) and not to the effective dynamics of
(\ref{eff}).  There should exist a separate argument showing that
(\ref{eff}) itself, which can also be viewed as a gauged $U(1)$ sigma
model in the unitary gauge, suffers from consistency issues of
\cite{Adams:2006sv}. We will now provide arguments showing that this
is indeed the case.

Consider specifically a model of the form which follows from (\ref{eff}) and (\ref{Vpot})
\be L = \int d^4 x\,  N \left[ -{1 \over 4} F_{\mu \nu} F^{\mu \nu} - {\lambda \over 4} (A_\mu A^\mu - v)^2 \right], \qquad \lambda > 0 , \qquad v < 0 \label{consider} \ . 
\ee

We will view this model as gauged $U(1)$ sigma model 
\be L = \int d^4 x\,  N \left[ -{1 \over 4} F_{\mu \nu} F^{\mu \nu} - {\lambda \over 4} ((\partial_\mu \sigma + A_\mu) (\partial^\mu \sigma + A^\mu) - v)^2 \right], \qquad \lambda > 0 , \qquad v < 0 \label{sigma} \ . 
\ee
In this formulation, it is manifest that (\ref{consider}) is the result of the $\sigma$ field being eaten by the gauge fields. The dynamics of this Goldstone boson before coupling to the gauge field has the form
\be  N \lambda \left( {v \over 2}  \partial_\mu \sigma \partial^\mu \sigma  - {1 \over 4} (\partial_\mu \sigma \partial^\mu \sigma)^2 \right), \qquad \lambda > 0 , \qquad v < 0 \ . \ee
Because $v$ is taken to be negative, this is precisely a model of the Ghost condensation type \cite{ArkaniHamed:2003uy}.  The $v$ and $\lambda$ are chosen so that $\partial \sigma$ is unstable to acquiring a space-like expectation value. Let us chose the vacuum  so that
\be \sigma = c x_3 + \tilde \sigma\ . \ee
As was explained in \cite{ArkaniHamed:2003uy}, there is a special
value $c = c^* = \sqrt{-v}$ which the model will flow to under Hubble
friction, but when decoupled from gravity, any $c > c^*$ will give
rise to a stable vacuum. The effective action will take the form
\beq L &=& N \lambda \left[-c (c^2+v)  \partial_3 \tilde \sigma + {(c^2 + v) \over 2} \partial_i \tilde \sigma \partial^i \tilde \sigma  - {(3 c^2 + v)\over 2} (\partial_3 \tilde \sigma)^2 + c \partial_3 \tilde \sigma (\partial_i \tilde \sigma \partial^i \tilde \sigma - (\partial_3 \tilde \sigma)^2)\right. \cr
&&\left. -{1 \over 4}  (\partial_i \tilde \sigma \partial^i \tilde \sigma - (\partial_3 \tilde \sigma)^2)^2\right] \label{sigma2} \eeq
where $i = 0,1,2$. The term which is first order in $\tilde \sigma$ is
a total derivative and can be ignored for this model. It is clear that
when one restricts to the sector where $\partial_3 \tilde \sigma=0$,
this model is precisely the prototype (\ref{prototype}) exhibiting the
obstruction to UV completion. This is already a strong hint that
(\ref{consider}) will exhibit similar pathology.

The remaining task is to show what happens to the dynamics of the
$\tilde \sigma$ field when it is eaten by the $U(1)$ gauge field. If
one chooses unitary gauge, this essentially amounts to replacing
$\partial_\mu \tilde \sigma$ by $A_\mu$ in (\ref{sigma2}). This time,
the linear term can not be ignored, and we are forced to set $c=c^* =
\sqrt{-v}$. With this choice, the effective action of the form
\beq L &=& N \lambda \left[-{1 \over 4 \lambda} F_{\mu \nu} F^{\mu \nu}  
 + v  A_3^2  \right. \cr
&&\left.
 + \sqrt{-v} A_3  (A_i  A^i  - A_3 ^2)
-{1 \over 4}  (A_i  A^i  - A_3^2)^2\right] \eeq
This effective action is manifestly invariant under the 2+1
dimensional Lorentz symmetry. 

In order to identify the manifestation of the UV obstruction, let us
consider the forward scattering of $A_i$ in the 2+1 Lorentz invariant
subspace, but with a small amount of Kaluza-Klein mass $\mu^2$ from
momentum along the $x_3$ direction. Let us take $\mu^2 \ll -\lambda v$
so that we can ignore the dynamics of the $A_3$ component. The
effective action can then be written in the form
\beq L &=& N \left[-{1 \over 4} F_{ij} F^{ij} +  {\mu^2 \over 2} A_i  A^i 
-{\lambda \over 4}  (A_i  A^i)^2\right] \ . \eeq

Now, let us chose to consider the forward scattering of the
longitudinal modes for a collision along the $x_1$ axis. The momentum
and the polarization of the ingoing particles will be set to
\be p_i = (E,p,0), \qquad \epsilon_i = {(p,E,0) \over \mu} \ee
and
\be p_i = (E,-p,0), \qquad \epsilon_i = {(-p,E,0) \over \mu} \ee
with $E^2 - p^2 = \mu^2$.
The forward scattering amplitude due to the $\lambda (A_i A^i)^2$ term will evaluate in this case to 
\be {\cal M} = - {\lambda \over N} {s^2 \over \mu^4} + {\cal O}(s) \ . \label{nutshell} \ee
We see that the $s^2$ term enters with a negative coefficient. This is
precisely the signature of the obstruction to UV completion described
in \cite{Adams:2006sv}.

The main point of this note, in a nutshell, can be reduced to the
arguments outlined between equations (\ref{consider}) and
(\ref{nutshell}). Since the conclusion relies only on the sign of
(\ref{nutshell}) which can be delicate to compute, we chose to include
preliminary arguments leading up to this conclusion. Note, in particular, that the signs of (\ref{higgs}) and (\ref{nutshell}) has the same origin.

It should be noted that the issues of UV completability have more to
do with the sign of $v_4$ rather than the sign of $v_2$.  Recall, in
the discussion below (\ref{mulambda2}), that the sign of $v_2$
dictated whether or not one expects $A_\mu$ to develop a vacuum
expectation value. To keep the action properly bounded, however, it
was necessary to keep $v_4$ positive.  This conclusion is more general
than the specific form of the potential (\ref{Vpot}) that we
considered. Any generic potential
\be V(A_\mu A^\mu) = V(A_i A^i - A_3^2) \ee
which is minimized at $A_i=0$, $A_3=\sqrt{-v}$ so that
\be V'(v) = 0, \qquad  V''(v) > 0 \ee
will contain a term
\be {1 \over 2} V''(v) (A_i A^i)^2 + {\cal O} (A_3) \ee
which enters the Lagrangian with a negative sign. 

It might be worth pointing out that in terms of the original variable
field $A_\mu$, the operator $v_4 (A_\mu A^\mu)^2$ whose coefficient we
identified as the diagnostic of the UV consistency following
\cite{Adams:2006sv} is non-derivative unlike for the case of the
scalar field in the example of (\ref{prototype}). This is not too
surprising in light of the fact that higher spin fields are in a
certain sense more non-local than the spinless fields.

In fact, it is also very unnatural for $v_2$ to take on a positive
value since it amounts to assigning a wrong sign for the kinetic term
of the Higgs field in (\ref{eff2}).  From the fermion model point of
view, this corresponds to $\lambda_2$ being positive, which is also
unnatural. Specifically, it can not arise from the exchange of massive
vector bosons. The $\lambda_2$ coupling is indeed bounded in the case
of the Thirring model. Although we did not find it in this article, it
seems very natural for such bound to follow from the consideration of
UV completability along the lines of \cite{Adams:2006sv} as well. This
will provide additional perspective on the dynamical treatment of the
fermion models
\cite{Bjorken:1963vg,Banks:1980rh,Jenkins:2003hw,Bjorken:2001pe}.

There are other examples of manifestly Lorentz invariant effective
field theories which supports superluminal backgrounds in non-trivial
backgrounds. One is the effective field theory of the ``$k$-essence''
type \cite{ArmendarizPicon:2000ah} whose Lagrangian takes the form
\be L = \int d^4 x\, f(\partial_\mu \phi \partial^\mu \phi) \ee
for some function $f$.\footnote{A special limit where $f(X) =
\sqrt{X}$, and where the speed of sound is infinite, is the
``cuscuton'' model of \cite{Afshordi:2006ad}.}\footnote{Another
possible pathology with the action of this form is that the map
between velocity and momentum field variables may not be one to one,
for example, if $f(X) = - a X + b X^2$ for $a,b>0$.}  Such an
effective action have long been known to support superluminal
propagation if $c_s^2 = (1+2X f''/f')^{-1} > 1$
\cite{Aharonov:1969vu,Garriga:1999vw,Babichev:2007dw}. This is
precisely the case for which $k$-essence models could have interesting
cosmological implications \cite{Bonvin:2006vc}. Another example is a
model of the form
\be 
  L = -\beta_1 \partial^\mu A^\sigma \partial_\mu A_\sigma - \beta_2
  (\partial_\mu A^\mu)^2 - \beta_3 \partial^\mu A^\sigma \partial_\sigma A_\mu
  + \lambda(A^{\mu} A_{\mu} -m^2)
\ee
which was considered in
\cite{Jacobson:2000xp,Jacobson:2004ts,Carroll:2004ai,Lim:2004js,Li:2007vz,Jacobson:2008aj}.
Here, $\lambda$ plays the role of the Lagrange multiplier constraining
$A^\mu A_\mu = m^2$, where the case of $m^2 >0$ corresponding to
time-like (therefore distinct from the space-like condensate
considered in \cite{Kraus:2002sa}) vacuum expectation value was
considered in \cite{Carroll:2004ai,Lim:2004js}, where it was shown
that there will be superluminal propagating mode if $(\beta_1+\beta_2
+ \beta_3)/\beta_1 > 1$.  These effective field theory models are
excellent candidates to look for concrete signatures obstructing
consistent UV completions, which we hope to identify in a future
work. A non-trivial part of this program might involve systematic
generalizations of the analysis of \cite{Adams:2006sv}.

We should comment before closing that the obstructions to UV
completion discussed in this note do not necessarily imply that such
effective field theories are not realized in nature.  The diagnostic
of \cite{Adams:2006sv} is mainly the probe of analyticity.  But nature
may well be non-analytic. It is still very interesting to study exotic
models and unconventional frameworks in the search for possible
solution to outstanding puzzles in effective field theory such as the
hierarchy problem and dark matter/dark energy. We just need to be
aware that some models may require sacrificing of analyticity.

Let us also note that there is one context in which gauge bosons
naturally arose as a consequence of spontaneous Lorentz symmetry
breaking. Embedding of flat D3-brane in type IIB string theory gives
rise to 6 massless scalar fields parameterizing the transverse
embedding of the brane. These are Goldstone bosons. For the D3-branes
in type IIB theory, however, there are enough supersymmetry to imply a
presence of massless spin particle in its multiplet. There may well
exist other, more elegant, ways to realize gauge fields as Goldstone
particles in spontaneous breaking of Lorentz symmetry.

\section*{Acknowledgements}

We would like to thank Daniel Chung and Ian Ellwood for discussions
and collaboration at early stage of this work. We also thank Per Kraus for extensive correspondences, and Alexander Vikman for comments. We also thank
Allan Adams, 
Baha Balenteken,
Jaques Distler, 
Charlie Goebel,
Tao Han,  and
Frank Petriello for discussions. This work was supported in
part by the DOE grant DE-FG02-95ER40896 and funds from the University
of Wisconsin.

\providecommand{\href}[2]{#2}\begingroup\raggedright\endgroup

\end{document}